\documentclass[pre,twocolumn]{revtex4-1}
\usepackage{hyperref}
\usepackage{amscd,amssymb,amsmath,latexsym,enumerate}
\usepackage[mathscr]{euscript}
\usepackage{epsfig}
\usepackage{fancybox}
\usepackage{verbatim}
\usepackage{graphicx,bm,units,yfonts,txfonts}

\newtheorem{theorem}{Theorem}
\newtheorem{corollary}{Corollary}

\newtheorem{proposition}{Proposition}
\newtheorem{remark}{Remark}

\newtheorem{example}{Example}

\newtheorem{code}{Computer Code}

\begin{document}

\title[Computational Many-Body Physics via $\mathcal M_{2^n}$ Algebra]{Computational Many-Body Physics via $\mathcal M_{2^q}$ Algebra}

\author{Emil Prodan}

\address{Department of Physics and Department of Mathematical Sciences, Yeshiva University, New York, USA}

\begin{abstract} 
The many-body Hamiltonians and other fermionic physical observables are expressed in terms of fermionic creation and annihilation operators, which, at an abstract level, form the algebra of Canonical Anti-commutation Relations. If the one-particle Hilbert space is $q$-dimensional, then this algebra is canonically isomorphic with the ordinary algebra $\mathcal M_{2^q}$ of $2^q \times 2^q$ matrices with complex entries. In this work, we present a method that makes this isomorphism explicit. This supplies concrete matrix representations of various many-body operators without involving the traditional Fock space representation. The result is a steep simplification of the many-body exact diagonalization codes, which is a significant step towards the soft-coding of generic fermionic Hamiltonians. Pseudo-code implementing matrix representations of various many-body operators are supplied and Hubbard-type Hamiltonians are worked out explicitly.
\end{abstract}

\thanks{Financial support through an award from W. M. Keck Foundation is acknowledged.}

\maketitle

\section{Introduction}

Local physical observables of fermionic systems are expressed as products and sums of creation $a_n^\ast$ and annihilation $a_n$ operators. The latter satisfy the canonical anti-commutation relations which automatically enforce Pauli's exclusion principle. The set of local fermionic physical observabales can be closed to and given the structure of a $C^\ast$-algebra, called the canonical anti-commutation relations algebra, or in short CAR-algebra \cite{BratelliBook2}. In the Heisenberg approach, one formulates the dynamics of fermions directly on the CAR-algebra and a many-body physical system is completely specified by a tuple $(\alpha,\mathcal T)$, where $\alpha$ is a group homomorphism $\alpha \, : \, \mathbb R \rightarrow {\rm Aut}({\rm CAR})$, specifying the time evolution of the physical observables, and $\mathcal T$ is a state invariant w.r.t. the $\alpha$-dynamics. In the Schroedinger picture, the dynamics of fermions is formulated on the anti-symmetric sector of the Fock space, which supplies a natural representation space for the CAR-algebra. Note that in Heisenberg's picture there is no place for Hilbert spaces and representations, and this observation is the starting point for our work. While we focus here at computational many-body aspects, this subtle but essential difference between the two pictures of quantum phenomena has conceptual consequences, as highlighted recently by Haldane in \cite{HaldaneJMP2018}.

At the computational level, the difference manifests as follows: In the Heisenberg picture, one seeks a direct homomorphism that embeds the algebra of observables in a matrix algebra. In Schroedinger's picture, the matrix representations are generated by acting with the operators on the basis of the Hilbert space. To see the major difference, let us consider a generic system where the fermions populate a discrete set $X$ of some physical space. We denote by $q$ the cardinal of $X$. The generic many-body observables take the form:
\begin{equation}\label{Eq:GenericElement}
A = \sum_{J,J' \subseteq X} a_{J,J'} \prod_{x \in J} a_x^\ast \prod_{x' \in J'} a_{x'}.
\end{equation}
Then in the Schroedinger picture, one will generate the $2^q \times 2^q$ matrix representation by looping over the occupation basis:
\begin{equation}
\langle n'_1 n'_2 \ldots n'_q|A|n_1 n_2 \ldots n_q\rangle, \quad n'_i,n_i \in \{0,1\},
\end{equation}
using, for example, the action of the generators on the basis:
\begin{align}
& a^\ast_m |n_1\ldots n_m \ldots n_q \rangle \\ \nonumber
& \qquad = (-1)^\alpha (n_m\oplus 1){\rm mod}\, 2 \, |n_1\ldots n_m+1 \ldots n_q \rangle,
\end{align}
where $\oplus$ is addition mod 2. In a successful Heisenberg program, however, one will use the embedding homomorphism to explicitly specify the entire $2^q \times 2^q$ matrix in one step. When the many-body operator has a simple structure, the action of $\prod_{x \in J} a_x^\ast \prod_{x' \in J'} a_{x'}$ on the occupation basis can be computed by hand and then the result can be integrated in the computer codes. However, this is typically hard-coded and the entire task needs to be repeated when presented with a different operator. A soft-code, by definition, is one that can diagonalize any many-body observable $A$ based on an input file that contains the subsets $J$ and $J'$ of $X$, as well as the associated coefficients $a_{J,J'}$. In the Schroedinger approach, the only way to achieve such soft-coding is to repeatedly apply the generators on the occupation basis but this leads to highly inefficient algorithms. This highlights one of the advantages of the Heisenberg approach, which become extremely useful when dealing with complicated Hamiltonians such as the ones often occuring in the research on topological phases of matters. For example, the Fidkowski-Kitaev  Hamiltonians \cite{FidkowskiPRB2010,FidkowskiPRB2011} contains products of as many as 8 generators! The model Hamiltonians for higher fractional Hall sequences \cite{ProdanPRB2009} present the same if not even higher level of complexity.

In this work, we exploit a well-know isomorphism between CAR and $M_{2^\infty}$ algebras \cite{DavidsonBook} to derive matrix representations of generic products of creation and annihilation operators. Explicit analytic formulas are supplied for several key products of generators, which will enable one to analytically translate any many-fermion Hamiltonian into a matrix form. For the reader's convenience, we exemplify the algorithms with concrete pieces of code and we work out several interesting many-fermion eigen-problems.

In our opinion, the benefits of the proposed approach can materialize in two extreme settings. The first one, is that of small-scale computations involving complex Hamiltonians. For example, the search and characterization of topological boundary modes in correlated systems require precisely this type of computations, especially when the goal is to validate their robustness against arbitrary interaction potentials. The challenge for this type of research is that the one-particle Hilbert spaces and the many-body Hamiltonians can vary drastically from one application to another and it is precisely this challenge that is addressed by our approach. The second setting is that of large-scale computations with standard two-body potentials, such as the Coulomb potential. Since our approach supplies formal matrix representations of the Hamiltonians, one can estimate the sparseness of the matrices (see for example Fig.~\ref{Fig:Sparce}) and then decide more easily on the optimal linear-algebra package to be used. One can also estimate more accurately the numerical errors and speed-up of the computations can result from the analytically determined action of the whole Hamiltonian on vectors.

\section{Background}

\subsection{The Algebra of Canonical Anti-Commutation Relations}

The algebra of canonical anti-commutation relations (CAR) is defined \cite{BratelliBook2} by a linear map $a:\mathcal H \rightarrow \mathcal B(\mathcal H')$ from a Hilbert space $\mathcal H$ onto the algebra of linear maps over another Hilbert space $\mathcal H'$, satisfying the following algebraic relations:
\begin{equation}\label{Eq:Rel1}
\begin{array}{c}
 a(f)a(g) + a(g)a(f)=0 \\
 a(f)^*a(g) + a(g) a(f)^* = (g,f)1.
\end{array}
\end{equation}
for all $f,\ g\in \mathcal H$. Here and throughout, $(\ ,\ )$ denotes the scalar product on $\mathcal H$. The CAR-algebra is the $C^*$-algebra generated by $\{a(f): f\in \mathcal H\}$ modulo relations~\eqref{Eq:Rel1}, endowed with the $\ast$-operation and the $C^*$-norm borrowed from $\mathcal B(\mathcal H')$. Up to an isomorphism, this definition is completely independent of the concrete representations of the Hilbert spaces. In many-body physics, $\mathcal H$ represents the one-particle Hilbert space and $\mathcal H'$ is chosen as the Fock-space and one says that $a(f)^\ast$ creates a fermion in the quantum state $f$, while $a(f)$ destroys a fermion in quantum state $f$. 

For condensed matter physicists, perhaps a more familiar representation of the CAR-algebra can be given in the following terms. Let $\{f_i\}_{i=\overline{0,\infty}}$ denote an orthonormal basis on $\mathcal H$ and let $a_i = a(f_i)$. Then the $a_i$'s satisfy the familiar anti-commutation relations:
\begin{align}\label{Eq:Rel2}
 a_i a_j + a_j a_i=a_i^\ast a_j^\ast + a_j^\ast a_i^\ast=0, \ a_i^\ast a_j + a_j a_i^\ast = \delta_{ij}\, 1.
\end{align}
If one prefers to maintain the liberty of choosing and changing the basis of the Hilbert space, the first representation in Eq.~\ref{Eq:Rel1} is definitely more preferable. 

 We will denote the CAR-algebra over a finite dimensional Hilbert space $\dim \mathcal H_q =q <\infty$ by $CAR(q)$. Throughout our presentation, we will be consistent and enumerate the elements of the orthonormal basis starting from 0 and ending at $q-1$. In other words, we will label the orthonormal basis of $\mathcal H_q$ as $f_0$, $f_1$, \ldots, $f_{q-1}$.
 
 The CAR-algebra is a $C^\ast$-algebra, that is, it is closed under the addition, multiplication and the $\ast$-transformation (or dagger-operation). The CAR-algebra also comes equipped with a norm but, since we are mainly considering finite CAR-algebras, this norm will not play any special role here. If $a(f), a(g),\ldots$ are some elements of $CAR(q)$, we will denote by $C^\ast \big( a(f),a(g),\ldots \big)$ the sub-algebra generated by them. Henceforth, $C^\ast \big( a(f),a(g),\ldots \big)$ contains all elements in $CAR(q)$ that can be formed through sums, multiplications and $\ast$-transformations of $a(f)$, $a(g)$, \ldots. In particular, let us point out that $CAR(q)$ can be naturally embedded in $CAR(q+1)$ and this sets an inductive tower which enable one to define $CAR(\mathcal H_\infty)$ as its inductive limit.

\subsection{The algebra $\mathcal M_{2^\infty}$}

Let $\mathcal M_2$ denote the algebra of $2\times 2$ matrices with complex entries. Then $\mathcal M_{2^q}=\mathcal M_2 \otimes \mathcal M_2 \ldots \otimes \mathcal M_2=\mathcal M_2^{\otimes q}$, which is isomorphic to the algebra of $2^q \times 2^q$ matrices with complex entries. Note that $\mathcal M_{2^q}$ can be embedded in $\mathcal M_{2^{q+1}}$ as $\left (^{\mathcal M_{2^q}}_0 \ ^0_{\mathcal M_{2^q}}\right )$ and, as such, one can set an inductive tower and define the UHF-algebra $\mathcal M_{2^\infty}$ as its inductive limit. The result is one of the most studied $C^\ast$-algebras in the mathematics literature. For example, its K-theory was worked out in \cite{RenaultBook} (see also \cite{DavidsonBook}).

We now introduce notations and conventions for our exposition. For $A\in \mathcal M_2$ we choose to write $A=\left (^{A_{00}}_{A_{10}} \ ^{A_{01}}_{A_{11}}\right )$. As a linear space, $\mathcal M_{2^q}$ is generated by the system of units 
\begin{equation}\label{GeneratorsM}
\big \{E^{(q)}_{mn} \big \}_{m,n=0,\ldots,2^{q}-1},
\end{equation}
 where $E^{(q)}_{mn}$ is the $2^{q}\times 2^{q}$ matrix with entry 1 at position $(m,n)$ and 0 in rest. The system of units satisfies the usual algebraic relations:
 \begin{equation}\label{Eq:M2qGen}
E^{(q)}_{mn} \, E^{(q)}_{m'n'} = \delta_{nm'} \, E^{(q)}_{mn'}.
\end{equation} 
The system of units for $\mathcal M_2$ will be denoted by $\{e_{\alpha \beta}\}_{\alpha,\beta=0,1}.$

\subsection{The link between the algebras}

\begin{theorem}\label{Link} $CAR(q)$ is isomorphic to $\mathcal M_{2^{q}}$ for all $q \in \mathbb N$.
\end{theorem}
{\bf \it Proof.} A detailed proof can be found in Kenneth Davidson's monograph \cite{DavidsonBook}. It will be, however, very instructive and helpful to present the proof in details once again here. Henceforth, let $f_0,f_1,\ldots, f_{q-1}$ be an orthonormal basis of $\mathcal H_q$ and set $a_i=a(f_i)$. Then $CAR(q)$ is simply $C^\ast\big(a_0,a_1,\ldots,a_{q-1}\big )$. Our first task is to define a new set of generator that commute with each other rather than anti-commute. This can be accomplished via a Jordan-Wigner type transformation, whose main mechanism is recalled below. 

Let $f$ be a normalized vector from $\mathcal H$ and let:
\begin{equation}
n_f \coloneqq a(f)^\ast a(f).
\end{equation}
 Since $a(f)^2=0$ and $a(f)^\ast a(f)+a(f)a(f)^\ast=1$, by multiplying the latter by $a(f)a(f)^\ast$, one obtains $n_f^2 = n_f$. In other words, $n_f$ is an idempotent for any norm-one vector $f$ from $\mathcal H_q$. In fact, $n_f$ is an orthogonal projector because $n_f^\ast =n_f$. Furthermore, 
\begin{equation}
a(f)a(f)^\ast=1-a(f)^\ast a(f) = 1 - e(f),
\end{equation}
hence $a(f)a(f)^\ast$ is the orthogonal complement of $n_f$:
\begin{equation}
n_f^\bot \coloneqq a(f)a(f)^\ast, \ n_f n_f^\bot=0, \ n_f + n_f^\bot = 1.
\end{equation}
Consider now another vector $g$ from $\mathcal H_q$ which is orthogonal on $f$, $(f,g)=0$. One can verify directly that $a(g)$ commutes with $n_f$ (hence also with $n_f^\bot$) but of course, $a(g)$ does not commute with $a(f)$. This can be fixed as follows. Define:
\begin{equation}
v \coloneqq n_f^\bot -n_f,
\end{equation}
with the following obvious properties:
\begin{equation}
v^*=v, \ v^2=1, \ v a(f)=-a(f) v, \ va(f)^* = -a(f)^* v.
\end{equation}
If $v a(g)$ is considered instead of $a(g)$, then:
\begin{align}
v a(g) a(f)=-v a(f)a(g) =  a(f)v a(g).
\end{align}
Similarly:
\begin{align}
v a(g) a(f)^*=-v a(f)^*a(g) =  a(f)^*v a(g).
\end{align}
Hence, the substitution $a(g) \rightarrow v a(g)$ made the operators commute. This is the essence of the Jordan-Wigner transformation. 

Returning now to $C^*\big(a_0,a_1,\ldots,a_{q-1}\big )$, we can define a set of commuting generators by iterating the above construction. This leads us to the following substitutions:
\begin{equation}
a_i \rightarrow v_i a_i,
\end{equation}
where $v_0 = 1$ and:
\begin{equation}
v_i = (n_{i-1}^\bot - n_{i-1}) v_{i-1}, \ v_i^\ast = v_i, \ v_i^2 = 1,  \ i=1,\ldots,q-1,
\end{equation}
with $n_i = a_i^\ast a_i$. It is important to keep in mind that $n_i$'s are all commuting orthogonal projections. The conclusion so far is that:
\begin{equation}
C^\ast\big(a_0,a_1,\ldots,a_{q-1}\big ) = C^\ast\big(v_0 a_0,v_1 a_1,\ldots,v_{q-1} a_{q-1}\big ),
\end{equation}
and now all the generators commute with each other. This concludes the step of the proof that involves the Jordan-Wigner transformation.

The next step is to look at the sub-algebra generated by each of these generators. Because of the anti-commution relations, one readily finds that $C^*\big ( v_i a_i \big )$ coincides with the $\mathbb C$-linear span of just four operators:
\begin{equation}
C^*\big ( v_i a_i \big )=\mathbb C\mathrm{-Span}\big \{n_i,n_i^\bot, v_i a_i, v_i a_i^\ast \big \}.
\end{equation}
Furthermore, if one sets:
\begin{equation}\label{Eq:es}
e^{(i)}_{11}=n_i, \ e^{(i)}_{00}=n_i^\bot, \ e^{(i)}_{01}=v_i a_i,\ e^{(i)}_{10}=v_i a_i^*,
\end{equation}
then
\begin{equation}\label{GeneratorsM2}
e^{(i)}_{\alpha \beta} e^{(i)}_{\alpha' \beta'} = \delta_{\beta \alpha'} e^{(i)}_{\alpha \beta'}
\end{equation}
which are exactly the algebraic relations satisfied by the generators of $\mathcal M_2$. Hence, Eq.~\ref{GeneratorsM2} defines an explicit isomorphic mapping of $\mathcal M_2$ into $C^*\big ( v_i a_i \big )$.

The last step of the proof involves the following elements of $CAR(q)$:
\begin{equation}\label{Eq:Es}
E^{(q)}_{\varphi,\psi} = e^{(0)}_{\varphi(0) \psi(0)} e^{(1)}_{\varphi(1) \psi(1)} \ldots e^{(q-1)}_{\varphi(q-1) \psi(q-1)},
\end{equation}
where $\varphi$ and $\psi$ are two functions of the type:
\begin{equation}
\varphi, \ \psi: \big \{0,1,\ldots,q-1 \big \} \rightarrow \{0,1\}.
\end{equation}
Note that there are exactly $2^{q}$ distinct such functions and one can verify explicitly that \eqref{Eq:Es} span the entire $CAR(q)$ as well as that:
\begin{equation}\label{Eq:CEs}
E^{(q)}_{\varphi,\psi} \ E^{(q)}_{\varphi' \psi'} = \delta_{\psi \varphi'} \ E^{(q)}_{\varphi \psi'},
\end{equation}
which are precisely the algebraic relations \eqref{Eq:M2qGen} defining the generators of $\mathcal M_{2^{q}}$. The conclusion is that:
\begin{equation}\label{Mapping1}
\left \{ E^{(q)}_{\varphi,\psi} : \varphi,\ \psi : \big \{0,\ldots,q-1 \big \} \rightarrow \big \{0,1 \big \} \right \} 
\end{equation}
supply an explicit isomorphic mapping of $\mathcal M_{2^{q}}$ into $CAR(q)$. Furthermore, this mapping respects the embedding of $CAR(q)$ into $CAR(q+1)$ and of $\mathcal M_{2^q}$ into $\mathcal M_{2^{q+1}}$, hence the inductive towers are isomorphic and their limits are isomorphic as $C^\ast$-algebras \cite{Effros1979}. $\square$
 
 \section{Practical Representations}
 
 For practical applications, we need to devise an efficient way to account for all $\varphi$'s and $\psi$'s appearing in Eq.~\ref{Mapping1}.
 
 \begin{proposition}
 Let $n$ be an integer between $0$ and $2^{q}-1$. Let:
 \begin{equation}
 n=\alpha_0 \cdot 2^0 + \alpha_1  \cdot 2^1 + \ldots \alpha_{q-1} \cdot 2^{q-1}, \quad \alpha_i \in \big \{0,1 \big \},
 \end{equation}
 be its unique binary representations and define: 
 \begin{equation}
 \mathfrak b_n: \big \{0,\ldots, q-1 \big \} \rightarrow \big \{0,1 \big \}, \ \mathfrak b_n(i)= \alpha_i,
 \end{equation}
 to be the function which outputs the binary digits of $n$. Then, when $n$ is varied from $0$ to $2^{q}-1$, the $\mathfrak b_n$'s generate all the possible functions $\varphi$'s and $\psi$'s appearing in Eq.~\ref{Mapping1}. 
 \end{proposition}
 
 \begin{remark}{\rm We introduce the following important conventions. Firstly, we will identify the elements $e^{(i)}_{\alpha\beta}$ of $CAR(q)$ introduced in \eqref{Eq:es} with the generators of $\mathcal M_2$ appearing at position $i$ in the tensor product $\mathcal M_2 \otimes \ldots \otimes \mathcal M_2$, tensored by the identity operators of the $\mathcal M_2$'s appearing at the other positions. Secondly, the system of units $E^{(q)}_{nm}$ generating $M_{2^{q}}$ and introduced in Eq.~\ref{GeneratorsM} will be identified with the elements of $CAR(q)$ via \eqref{Eq:CEs}:
 \begin{equation}
 E^{(q)}_{nm} := E^{(q)}_{\mathfrak b_n \mathfrak b_m},
 \end{equation}
 and, as such, we will use the notations interchangeably.
} $\Diamond$
 \end{remark}
 
 The above proposition and Theorem~\ref{Link} provides the following important Corollary.
 
 \begin{corollary}\label{MainRepresentation}
 Let $\alpha_0 \alpha_1 \ldots \alpha_{q-1}$ and $\beta_0 \beta_1 \ldots \beta_{q-1}$ be two binary sequences of $1$'s and $0$'s. Then:
 \begin{equation}
e^{(0)}_{\alpha_0 \beta_0} e^{(1)}_{\alpha_1 \beta_1} \ldots e^{(q)}_{\alpha_{q-1} \beta_{q-1}}=E^{(q)}_{nm},
\end{equation}
where
\begin{align}
& n=\alpha_0 \cdot 2^0 + \alpha_1  \cdot 2^1 + \ldots \alpha_{q-1} \cdot 2^{q-1},\\
& m=\beta_0 \cdot 2^0 + \beta_1  \cdot 2^1 + \ldots \beta_{q-1} \cdot 2^{q-1}.
\end{align}
Conversely, for any $m$ and $n$ between $0$ and $2^{q}-1$ one has:
\begin{equation}
E^{(q)}_{nm} = 
 e^{(0)}_{\mathfrak b_n(0), \mathfrak b_m(0)} e^{(1)}_{\mathfrak b_n(1), \mathfrak b_m(1)} \ldots e^{(q)}_{\mathfrak b_n(q-1), \mathfrak b_m(q-1)}.
 \end{equation}
 \end{corollary}
 
\begin{code}{\rm Below are code lines which performs the binary decomposition of an integer number $n \in \{0,1,\ldots, 2^{q}-1\}$.
\begin{equation}\label{Binary}
\boxed{
\begin{array}{l}
 \ \ \mathrm{input} \ n \\
\ \  \mathfrak b_n(j)=0, \ \ j=0,\ldots, q-1 \\
 \ \ p=0 \\
      \ \   \mathrm{do}  j=0,q-1 \\
     \  \   \ \  \mathfrak b_n(j)=(n-p)/2^j \mod 2 \\
 \  \  \ \   p = p+\mathfrak b_n(j)*2^j \\
     \ \   \mathrm{end \ do} \\ 
     \ \ \mathrm{return} \ \mathfrak b_n(j), \ \ j=0,\ldots, q-1.
\end{array}
}
\end{equation}
} 
\end{code}
 
 \begin{example} {\rm Let us compute $e^{(0)}_{12} e^{(1)}_{22} e^{(2)}_{12}$ from $CAR(3) \simeq \mathcal M_{8}$. We have successively:
 \begin{align}
 e^{(0)}_{12} e^{(1)}_{22} e^{(2)}_{12} &
 = \left [
 \begin{array}{cc}
 0  & e^{(1)}_{22} e^{(2)}_{12}  \\
 0  & 0
 \end{array}
 \right] \\ \nonumber 
& = \left [
 \begin{array}{cccc}
 0 & 0 & 0 & 0 \\
 0 & 0 & 0 & e^{(2)}_{12} \\
 0 & 0 & 0 & 0 \\
  0 & 0 & 0 & 0 \\ 
 \end{array}
 \right ] 
 =\left [
 \begin{array}{cccc}
 0 & 0 & 0 & 0 \\
 0 & 0 & 0 & \boxed{^0_0 \ ^1_0} \\
 0 & 0 & 0 & 0 \\
  0 & 0 & 0 & 0 \\ 
 \end{array}
 \right ] 
 \end{align}
Above, all the $0$'s represent the null $2\times2$ matrix, excepting the $0$'s in the box, which are just ordinary $0$'s. On the other hand:
\begin{equation}
\begin{array}{l}
n=0\times 2^0+1\times 2^1+0\times 2^2=2, \\ 
m=1\times 2^0+1\times 2^1+1\times 2^2=7,
\end{array}
\end{equation}
hence Corollary~\ref{MainRepresentation} predicts:
\begin{equation}
e^{(0)}_{12} e^{(1)}_{22} e^{(2)}_{12}=E^{(3)}_{27},
\end{equation}
which is indeed the case (recall that we run the indices from 0 to 7).} $\Diamond$
\end{example}

\section{Matrix representations of many-fermion operators}

\subsection{Matrix representations of the generators}

As a model calculation, we derive first the matrix representations of $a_i$ and $a_i^\ast$ in $\mathcal M_{2^{q}}$. We start from:
\begin{equation}
a_i=v_i e^{(i)}_{01}=(n_0^\bot -n_0) \ldots (n_{i-1}^\bot -n_{i-1}) e^{(i)}_{01}.
\end{equation}
Using the definitions in Eq.~\ref{Eq:es} we obtain:
\begin{equation}\label{X1}
a_i= \left ( e^{(0)}_{00} - e^{(0)}_{11} \right ) \ldots \left ( e^{(i-1)}_{00} - e^{(i-1)}_{11} \right )e^{(i)}_{01}.
\end{equation}
Corollary~\ref{MainRepresentation} gives matrix representations for products of $e$'s that contain exactly $q$ terms. As such, we need to insert identity operators in Eq.~\ref{X1} until we complete the products:
\begin{align}
a_i=& \left ( e^{(0)}_{00} - e^{(0)}_{11} \right ) \ldots \left ( e^{(i-1)}_{00} - e^{(i-1)}_{11} \right )e^{(i)}_{01} \\ \nonumber 
& \qquad \qquad \times \left ( e^{(i+1)}_{00} + e^{(i+1)}_{11} \right ) \ldots \left ( e^{(q-1)}_{00} + e^{(q-1)}_{11} \right ).
\end{align}
Expanding:
\begin{equation}
a_i = \sum_{\alpha's} (-1)^{\alpha_0 +\ldots + \alpha_{i-1}} e^{(0)}_{\alpha_0 \alpha_0} \ldots e^{(i-1)}_{\alpha_{i-1}\alpha_{i-1}} e^{(i)}_{01} e^{(i+1)}_{\alpha_{i+1} \alpha_{i+1}} \ldots e^{(q-1)}_{\alpha_{q-1} \alpha_{q-1}}.
\end{equation}
The above sum is over the set of all binary sequences of the form $$\alpha_0 \alpha_1 \ldots \alpha_{i-1} 0 \alpha_{i+1} \ldots \alpha_{q-1},$$ which coincides with the set of the binary expansions of $n \in \{0,\ldots,2^{q}-1\}$ with $\mathfrak b_{n}(i)=0$. Using Corollary~\ref{MainRepresentation} and accounting for $\alpha$'s properly, we obtain a closed-form formula for $a_i$ and, by applying the $\ast$-operation, we also get a closed-form formula for $a_i^\ast$:

\begin{proposition}\label{Pro:As} In terms of the standard generators of $M_{2^{q}}$, we have:
\begin{align}
& a_i = \sum_{n=0}^{2^{q}-1} (-1)^{\sum_{s=0}^{i}\mathfrak b_n(s)} \ \delta_{\mathfrak b_n(i),0}\   E^{(q)}_{n,n+2^i},  \label{Formula1}\\
& a_i^\ast = \sum_{n=0}^{2^{q}-1} (-1)^{\sum_{s=0}^{i}\mathfrak b_n(s)} \ \delta_{\mathfrak b_n(i),0}\   E^{(q)}_{n+2^i,n}. \label{Formula2}
\end{align}
\end{proposition}

\begin{remark}{\rm We have verified analytically that the above matrices indeed satisfy the commutation relations \eqref{Eq:Rel2}.} $\Diamond$
\end{remark} 

\subsection{Matrix representations of products of generators}

We continue our computations with a derivation of the product $a_i^\ast a_j$, assuming for the beginning that $j > i$. Starting from \eqref{X1}, we have:
\begin{align}\label{Eq:Z1}
a_i^\ast a_j & = \left ( e^{(0)}_{00} - e^{(0)}_{11} \right ) \ldots \left ( e^{(i-1)}_{00} - e^{(i-1)}_{11} \right )e^{(i)}_{10} \\ \nonumber
& \quad \quad \times \left ( e^{(0)}_{00} - e^{(0)}_{11} \right ) \ldots \left ( e^{(j-1)}_{00} - e^{(j-1)}_{11} \right )e^{(j)}_{01}  \\ \nonumber
& = \left ( e^{(0)}_{00} + e^{(0)}_{11} \right ) \ldots \left ( e^{(i-1)}_{00} + e^{(i-1)}_{11} \right ) e^{(i)}_{10}\\ \nonumber
& \quad \quad  \times \left ( e^{(i+1)}_{00} - e^{(i+1)}_{11} \right ) \ldots \left ( e^{(j-1)}_{00} - e^{(j-1)}_{11} \right ) \\ \nonumber
&  \quad \quad \quad \times e^{(j)}_{01} \left ( e^{(j+1)}_{00} + e^{(j+1)}_{11} \right ) \ldots \left ( e^{(q-1)}_{00} + e^{(q-1)}_{11} \right ),
\end{align}
where the middle line is missing if $j=i+1$. Let us note that the case $i>j$ follows from the case treated above by applying the conjugation. Furthermore, we can straightforwardly modify the above arguments to find that, for $i<j$:
\begin{align}
a_i a_j^\ast &  = - \left ( e^{(0)}_{00} + e^{(0)}_{11} \right ) \ldots \left ( e^{(i-1)}_{00} + e^{(i-1)}_{11} \right ) e^{(i)}_{01}\\ \nonumber
& \quad \quad  \times \left ( e^{(i+1)}_{00} - e^{(i+1)}_{11} \right ) \ldots \left ( e^{(j-1)}_{00} - e^{(j-1)}_{11} \right ) \\ \nonumber
&  \quad \quad \quad \times e^{(j)}_{10} \left ( e^{(j+1)}_{00} + e^{(j+1)}_{11} \right ) \ldots \left ( e^{(q-1)}_{00} + e^{(q-1)}_{11} \right ),
\end{align}
and the reason for the minus sign is $e^{(i)}_{01}\left ( e^{(i)}_{00} - e^{(i)}_{11} \right ) = - e^{(i)}_{01}$, as opposed to $e^{(i)}_{10}\left ( e^{(i)}_{00} - e^{(i)}_{11} \right ) = + e^{(i)}_{10}$. After expanding and using Corollary~\ref{MainRepresentation}, we obtained:

\begin{proposition}\label{Pro:AA1} In terms of the standard generators of $M_{2^{q}}$, we have for $i \neq j$:
\begin{align}\label{Eq:AA11}
a_i^\ast a_j = & \sum_{n=0}^{2^{q}-1} (-1)^{\sum_{s={\rm min}(i,j)}^{{\rm max}(i,j)} \mathfrak b_n(s)} \  \delta_{\mathfrak b_n(i),0} \ \delta_{\mathfrak b_n(j),0} \ E^{(q)}_{n+2^i,n+2^j} \\ \nonumber 
= & - \sum_{n=0}^{2^{q}-1} (-1)^{\sum_{s={\rm min}(i,j)}^{{\rm max}(i,j)} \mathfrak b_n(s)} \  \delta_{\mathfrak b_n(i),1} \ \delta_{\mathfrak b_n(j),0} \ E^{(q)}_{n,n - 2^i + 2^j},
\end{align}
and $a_j a_i^\ast = - a_i^\ast a_j$. 
\end{proposition}

If $i=j$, the calculations gives:
\begin{align}
a_i^\ast a_i = & \left ( e^{(0)}_{00} + e^{(0)}_{11} \right ) \ldots \left ( e^{(i-1)}_{00} + e^{(i-1)}_{11} \right ) e^{(i)}_{11} \\ \nonumber 
& \qquad \qquad \times \left ( e^{(j+1)}_{00} + e^{(j+1)}_{11} \right ) \ldots \left ( e^{(q-1)}_{00} + e^{(q-1)}_{11} \right ),
\end{align}
and:
\begin{align}
a_i a_i^\ast  = & \left ( e^{(0)}_{00} + e^{(0)}_{11} \right ) \ldots \left ( e^{(i-1)}_{00} + e^{(i-1)}_{11} \right ) e^{(i)}_{00} \\ \nonumber 
& \qquad \qquad \times \left ( e^{(j+1)}_{00} + e^{(j+1)}_{11} \right ) \ldots \left ( e^{(q-1)}_{00} + e^{(q-1)}_{11} \right ).
\end{align}
The conclusion is:

\begin{proposition}\label{Pro:AA2} In terms of the standard generators of $M_{2^{q}}$, we have:
\begin{equation}\label{Eq:AA20}
n_i = a_i^\ast a_i = \sum_{n=0}^{2^{q}-1} \delta_{\mathfrak b_n(i),1} \ E^{(q)}_{n,n},
\end{equation}
\begin{equation}
n_i^\bot = a_i a_i^\ast = \sum_{n=0}^{2^{q}-1} \delta_{\mathfrak b_n(i),0} \ E^{(q)}_{n,n},
\end{equation}
\begin{equation}\label{Eq:AA21}
n_{i_1} n_{i_2} \ldots n_{i_k} = \sum_{n=0}^{2^{q}-1} \delta_{\mathfrak b_n(i_1),1} \ldots \delta_{\mathfrak b_n(i_k),1} \ E^{(q)}_{n,n },
\end{equation}
\begin{equation}\label{Eq:AA22}
n_{i_1}^\bot n_{i_2}^\bot \ldots n_{i_k}^\bot = \sum_{n=0}^{2^{q}-1} \delta_{\mathfrak b_n(i_1),0} \ldots \delta_{\mathfrak b_n(i_k),0} \ E^{(q)}_{n,n }.
\end{equation}
\end{proposition}

The products $a_i a_j$ and $a_i^\ast a_j^\ast$ can be treated similarly. 

\begin{proposition}\label{Pro:AA3} In terms of the standard generators of $M_{2^{q}}$, we have:
\begin{align}
a_i a_j = & \, {\rm sgn}(i-j)\sum_{n=0}^{2^{q}-1}   (-1)^{\sum_{s={\rm min}(i,j)}^{{\rm max}(i,j)} \mathfrak b_n(s)} \\ \nonumber
& \qquad  \qquad \qquad \times \ \delta_{\mathfrak b_n(i),0} \ \delta_{\mathfrak b_n(j),0} \ E^{(q)}_{n,n+2^i+2^j},
\end{align}
and:
\begin{align}
a_i^\ast a_j^\ast = & \, {\rm sgn}(j-i) \sum_{n=0}^{2^{q}-1}   (-1)^{\sum_{s={\rm min}(i,j)}^{{\rm max}(i,j)} \mathfrak b_n(s)} \\ \nonumber 
& \qquad \qquad \qquad \times \  \delta_{\mathfrak b_n(i),0} \ \delta_{\mathfrak b_n(j),0} \ E^{(q)}_{n+2^i+2^j,n},
\end{align}
where we adopt the convention that ${\rm sgn}(0)=0$.
\end{proposition}

\begin{remark}{\rm It will be convenient to introduce the notation:
\begin{equation}
\mathcal N_{ij}(n) = \sum_{s={\rm min}(i,j)}^{{\rm max}(i,j)} \mathfrak b_n(s), \quad i \neq j,
\end{equation}
since the sign factors determined by these coefficients will appear often in the subsequent presentation.} $\Diamond$
\end{remark}

A direct consequence of Proposition~\ref{Pro:AA3} is the following useful identity:

\begin{corollary} In terms of the standard generators of $M_{2^{q}}$, we have:
\begin{align}
a_i^\ast a_j^\ast a_k a_l= & \, {\rm sgn}\big [(j-i)(k-l) \big ] \sum_{n=0}^{2^{q}-1}  (-1)^{\mathcal N_{ij}(n)+\mathcal N_{kl}(n)} \\ \nonumber 
& \qquad \times \  \delta_{\mathfrak b_n(i),0} \, \delta_{\mathfrak b_n(j),0} \, \delta_{\mathfrak b_n(k),0} \, \delta_{\mathfrak b_n(l),0} \ E^{(q)}_{n+2^i+2^j,n+2^k+2^l} \\ \nonumber 
= & {\rm sgn}\big [(j-i)(k-l) \big ] \, \sum_{n=0}^{2^{q}-1} (-1)^{\mathcal N_{ij}(n)+\mathcal N_{kl}(n)} \\ \nonumber 
& \qquad \times \  \delta_{\mathfrak b_n(i),1} \, \delta_{\mathfrak b_n(j),1} \, \delta_{\mathfrak b_n(k),0} \, \delta_{\mathfrak b_n(l),0} \ E^{(q)}_{n,n-2^i - 2^j + 2^k+2^l}.
\end{align}
\end{corollary}

\begin{example} \label{Hubbard} {\rm We derive the matrix representation of the following Hubbard-type Hamiltonian:
\begin{equation}\label{Eq:Hub1}
H=\sum_{i,j=0}^{q-1} \Big (\delta_{ij} \, \epsilon_i \, n_i + (1 -\delta_{ij})\big (t_{ij} \ a_i^\ast \, a_{j} + \bar t_{ij} \ a_{j}^\ast \, a_i \big ) + u_{ij} \ n_i \, n_{j} \Big ),
\end{equation} 
where $t_{ij}$'s and $u_{ij}$'s and $\epsilon_i$'s are some complex and real parameters, respectively. Browsing through the list of formulas supplied above, one can see that the matrix representation of $H$ can be obtained automatically from Eqs.~\eqref{Eq:AA11}, \eqref{Eq:AA20} and \eqref{Eq:AA21}:
\begin{align}\label{Eq:H0}
H= & \sum_{n=0}^{2^{q}-1} \sum_{i,j=0}^{q-1}  \Big [ (-1)^{\mathcal N_{ij}(n)} \ \delta_{\mathfrak b_n(i),0} \ \delta_{\mathfrak b_n(j),0} \, (1 -\delta_{ij})  \\ \nonumber
& \qquad \qquad \times \Big ( t_{ij} \ E^{(q)}_{n+2^i,n + 2^{j}} + \bar t_{ij} \ E^{(q)}_{n+2^{j},n + 2^i} \Big )   \\ \nonumber
& \qquad \qquad \qquad +  \delta_{\mathfrak b_n(i),1} \,  \delta_{\mathfrak b_n(j),1}(u_{ij} +\delta_{ij} \epsilon_i) \, E^{(q)}_{n,n}\Big ]. \quad \Diamond
\end{align}}
\end{example}

\begin{code} {\rm We provide here a basic piece of code which computes and stores the entire matrix of $H$ from \eqref{Eq:H0} in $\mathcal M_{2^{q+1}}$.
\begin{equation}
\boxed{
\begin{array}{l}
h_{n,n'}=0, \ n,n'=0,\ldots, 2^{q}-1 \\
\mathrm{do} \ n=0,2^{q}-1 \\
\ \  \mbox{Call Eq.~\ref{Binary}} \\
\ \   \mathrm{do} \ i=0,q-1 \\
\ \ \ \  \mathrm{do} \ j=0,q-1 \\
\ \ \ \ \ \  \mathrm{if}(\mathfrak b_n(i) = \mathfrak b_n(j)= 0)  \ \mathrm{then} \\
\ \ \ \ \ \ \ \ \mathcal N_{ij}=\mathrm{sum}\big [\mathfrak b_n \big(\min(i,j) \, : \, \max(i,j)\big )\big ] \\ 
\ \ \ \ \ \ \ \ h_{n+2^i,n+2^{j}}=(-1)^{\mathcal N_{ij}} t_{ij} \\
 \ \ \ \ \ \ \ \ h_{n+2^{j},n+2^{i}}=(-1)^{\mathcal N_{ij}} \bar t_{ij} \\
\ \ \ \ \ \ \mathrm{end \ if}\\
\ \ \ \  \ \  \mathrm{if}(\mathfrak b_n(i)=\mathfrak b_n(j)= 1) \ \mathrm{then} \\  
\ \ \ \ \ \ \ \ h_{n,n}=\delta_{ij} \epsilon_i + u_{ij} \\
\ \ \ \ \ \ \mathrm{end \ if}\\
\ \ \ \ {\rm end \ do}
\ \  \mathrm{end \ do} \\
\mathrm{end \ do} \\
{\rm return} \ h_{n,n'}, \ n,n'=0, \ldots, 2^{q}-1.
\end{array}
}
\end{equation}
Let us highlight the simplicity of the code.} $\Diamond$
\end{code}

\begin{remark}{\rm Even though $H$ conserves the number of particles, an issue to be addressed in the next section, there are cases where computing the full matrix of $\hat H$ is still desirable, such as when $H$ is perturbed with a potential that does not conserves the number of particles.} $\Diamond$
\end{remark}

\section{N-particles sectors}

Our first goal is to give the spectral decomposition of the number of particles operator inside the algebra $\mathcal M_{2^{q}}$. We will then use its spectral sub-spaces to decompose the Hamiltonians in block diagonals.

\subsection{Spectral resolution of the particle number operator}

Let: 
\begin{equation}
\hat N=\sum_{i=0}^{q-1} a_i^\ast a_i = \sum_{i=0}^{q-1} n_i = \sum_{i=0}^{q-1} e_{11}^{(i)}
\end{equation}
be the classical particle-number operator. A direct way to generate its spectral decomposition inside $\mathcal M_{2^q}$ will be to complete $e_{11}^{(i)}$'s to full product sequences and follow the steps above. We, however, proceed slightly differently.

\begin{proposition} Let $n$ be a number between $0$ and $2^{q}-1$. Then:
\begin{equation}
\hat N \, E^{(q)}_{n,n} = E^{(q)}_{n,n} \, \hat N =\mathcal N(n) E^{(q)}_{n,n}, \quad \mathcal N(n) = \sum_{s=0}^{q-1} \mathfrak b_n(s).
\end{equation}
\end{proposition}
{\bf \it Proof.} From Corollary~\ref{MainRepresentation}:
\begin{equation}
E^{(q)}_{n,n} =  e^{(0)}_{\mathfrak b_n(0), \mathfrak b_n(0)} e^{(1)}_{\mathfrak b_n(1), \mathfrak b_n(1)} \ldots e^{(q-1)}_{\mathfrak b_n(q-1), \mathfrak b_n(q-1)}.
 \end{equation}
Since above all the $e$'s commute, we can separate the terms with $\mathfrak b_n(i)=1$ to the left and the remaining terms with $\mathfrak b_n(i)=0$ to the right. In this way, we obtain:
\begin{equation}
E^{(q)}_{n,n} = \prod_{\mathfrak b_n(i)=1} n_i \prod_{\mathfrak b_n(j)=0} n_j^\bot.
\end{equation}
Then:
\begin{align}
\hat N E^{(q)}_{n,n} & = \left (\sum_{k=0}^{q-1} n_k \right ) \prod_{\mathfrak b_n(i)=1} n_i \prod_{\mathfrak b_n(j)=0} n_j^\bot \\ \nonumber
& = \sum_{k=0}^{q-1} \big (\delta_{\mathfrak b_n(k),0} + \delta_{\mathfrak b_n(k),1} \big ) \, n_k\,  \prod_{\mathfrak b_n(i)=1} n_i \prod_{\mathfrak b_n(j)=0} n_j^\bot,
\end{align}
and since the $n_i$'s are projections, the last line can be written as:
\begin{align}
& \sum_{k=0}^{q-1} \delta_{\mathfrak b_n(k),1}  \prod_{\mathfrak b_n(i)=1} n_i \prod_{\mathfrak b_n(j)=0} n_j^\bot  \\ \nonumber 
& \qquad \qquad = \left (\sum_{k=0}^{q-1} \mathfrak b_n(k) \right ) \prod_{\mathfrak b_n(i)=1} n_i \prod_{\mathfrak b_n(j)=0} n_j^\bot
\end{align}
and the statement follows. $\square$

\begin{corollary} The spectral decomposition of $\hat N$ is:
\begin{equation}
\hat N = \sum_{N=0}^{q-1}  N \ \Big ( \sum_{n=0}^{2^q-1} \delta_{\mathcal N(n),N} \ E^{(q)}_{n,n} \Big ).
\end{equation}
\end{corollary}
{\it Proof.} The family of rank-one projections $E^{(q)}_{n,n}$, $n=0,2^{q}-1$ gives a resolution of the identity in $\mathcal M_{2^{q}}$:
\begin{equation}
\sum_{n=0}^{2^{q}-1} E^{(q)}_{n,n} = I_{2^{q} \times 2^{q}}.
\end{equation} 
Hence, the rangel of the projections $E^{(q)}_{n,n}$ exhaust all the invariant Hilbert sub-spaces of $\hat N$ when $n$ is varied from $0$ to $2^{q}-1$, and the statement follows. $\square$.

\begin{code}{\rm We provide below lines of code that detect and re-label the original indices that belong to a specific $N$-particle sector. We call these new indices the $N$-compressed indices.
\begin{equation}\label{Eq:IndN}
\boxed{
\begin{array}{l}
\mathrm{input} \ N \\
\mathrm{ind}(n)=0, \ n=0,\ldots,2^q-1, \\
c=0 \\
\mathrm{do} \ n=0,2^q-1 \\
\ \  \mbox{Call Eq.~\ref{Binary}} \\
\ \  \mathcal N(n)=\mathrm{sum}(\mathfrak b_n) \\
\ \ \mathrm{if}(\mathcal N(n) =N) \ \mathrm{then} \\
\ \ \ \ c=c+1 \\
\ \ \ \ \mathrm{ind}(n)=c \\
\ \   \mathrm{end \ if} \\
\mathrm{enddo}\\
D_N = c \\
{\rm return} \ D_N, \ {\rm ind}(n), \ n = 0,\ldots,2^q-1.
\end{array}
}
\end{equation}
These new indices will be used to generate, store and manipulate the diagonal blocks of the Hamiltonians corresponding to the $N$-particle sectors. Note that $D_N$ is the dimension of the N-particle sector.} $\Diamond$
\end{code}

\subsection{Elementary operators on N-particle sectors}

Let $\Phi(a)$ be a product of $a$'s with equal number of creation and annihilation generators. Then $\Phi(a)$ commutes with $N$ and the $N$-th block of the product can be computed from:
\begin{align}
\Phi(a)_N = & \, \Big ( \sum_{n=0}^{2^q-1} \delta_{\mathcal N(n),N} \ E^{(q)}_{n,n} \Big ) \ \Phi(a) \\ \nonumber 
& = \Phi(a) \ \Big ( \sum_{n=0}^{2^q-1} \delta_{\mathcal N(n),N} \ E^{(q)}_{n,n} \Big ).
\end{align}
Applying this procedure on the products in Propositions~\ref{Pro:AA1} and \ref{Pro:AA2} gives:

\begin{proposition}\label{Pro:NAA} In terms of the standard generators of $M_{2^{q}}$, we have:
\begin{align}
& \big (a_i^\ast a_j \big )_N = -\big ( a_j a_i^\ast \big )_N = \\ \nonumber
&  \quad - \sum_{n=0}^{2^{q}-1} (-1)^{\mathcal N_{ij}(n)} \  \delta_{\mathcal N(n),N}\, \delta_{\mathfrak b_n(i),1} \ \delta_{\mathfrak b_n(j),0} \ E^{(q)}_{n,n - 2^i + 2^j},
\end{align}
\begin{equation}
\big (n_{i_1} n_{i_2} \ldots n_{i_k} \big )_N = \sum_{n=0}^{2^{q}-1} \delta_{\mathcal N(n),N} \, \delta_{\mathfrak b_n(i_1),1} \ldots \delta_{\mathfrak b_n(i_k),1} \ E^{(q)}_{n,n }.
\end{equation}
\begin{equation}
\big (n_{i_1}^\bot n_{i_2}^\bot \ldots n_{i_k}^\bot \big )_N= \sum_{n=0}^{2^{q+1}-1} \delta_{\mathcal N(n),N} \, \delta_{\mathfrak b_n(i_1),0} \ldots \delta_{\mathfrak b_n(i_k),0} \ E^{(q)}_{n,n }.
\end{equation}
\begin{align}
& \big ( a_i^\ast a_j^\ast a_k a_l \big )_N= {\rm sgn}\big [(j-i)(k-l) \big ]  \sum_{n=0}^{2^{q}-1} \ (-1)^{\mathcal N_{ij}(n)+\mathcal N_{kl}(n)} \\ \nonumber 
& \quad \times \ \delta_{\mathcal N(n),N}\, \delta_{\mathfrak b_n(i),1} \, \delta_{\mathfrak b_n(j),1} \, \delta_{\mathfrak b_n(k),0} \, \delta_{\mathfrak b_n(l),0} \ E^{(q)}_{n,n-2^i-2^j+2^k+2^l}.
\end{align}
\end{proposition}

The particle number operator commutes with any product of generators which contains an equal number of creation and annihilation operators. In particular $N$ commutes with the Hamiltonian defined in Example~\ref{Hubbard}. Its block diagonals are worked out below.

\begin{example}\label{HubbardN} {\rm In the $N$-particle sector, the Hubbard model from Example~\ref{Hubbard} becomes:
\begin{align}\label{Eq:HHHN}
H_N = & \sum_{i,j=0}^{q-1} \Big ( (1 -\delta_{ij})\big (t_{ij} \ \big (a_i^\ast \, a_{j}\big )_N + \bar t_{ij} \ \big (a_{j}^\ast \, a_i \big)_N \big ) \\ \nonumber 
& \qquad \qquad + (\delta_{ij} \, \epsilon_i+u_{ij}) \big( \ n_i \, n_j \big )_N \Big ),
\end{align} 
and its matrix form can be automatically generated from Proposition~\ref{Pro:NAA}:
\begin{align}\label{Eq:HN}
H_N = & \sum_{n=0}^{2^q-1} \delta_{\mathcal N(n),N}\sum_{i,j=0}^{q-1}   \Big [(-1)^{\mathcal N_{ij}(n)+1} \, \delta_{\mathfrak b_n(i),1} \, \delta_{\mathfrak b_n(j),0} \, (1 -\delta_{ij})   \\ \nonumber 
& \qquad \qquad \qquad \times \Big ( t_{ij} \,  \, E^{(q)}_{n,n - 2^i+ 2^j} +  \bar t_{ij} \, \, E^{(q)}_{n-2^i + 2^j,n } \Big )  \\ \nonumber 
& \qquad \qquad \qquad \qquad +  \delta_{\mathfrak b_n(i),1} \, \delta_{\mathfrak b_n(j),1} \,  (\delta_{ij} \epsilon_i + u_{ij}) \ E^{(q)}_{n,n}\Big ]. \ \Diamond
\end{align}
}
\end{example}

\begin{remark}{\rm Comparing with Eq.~\eqref{Eq:H0}, we see that the only change in \eqref{Eq:HN} is a selective summation over $n$. However, when resolving over the particle number sectors, the computational challenge  is two-fold: (a) determining the reduced form of the Hamiltonian, which \eqref{Eq:HN} delivers, and (b) storing this reduced Hamiltonian using a minimal and natural set of indices. It is at this point where the indices introduced in \eqref{Eq:IndN} become useful, as we will see below.
} $\Diamond$
\end{remark}

\begin{figure}
\includegraphics[width=\linewidth]{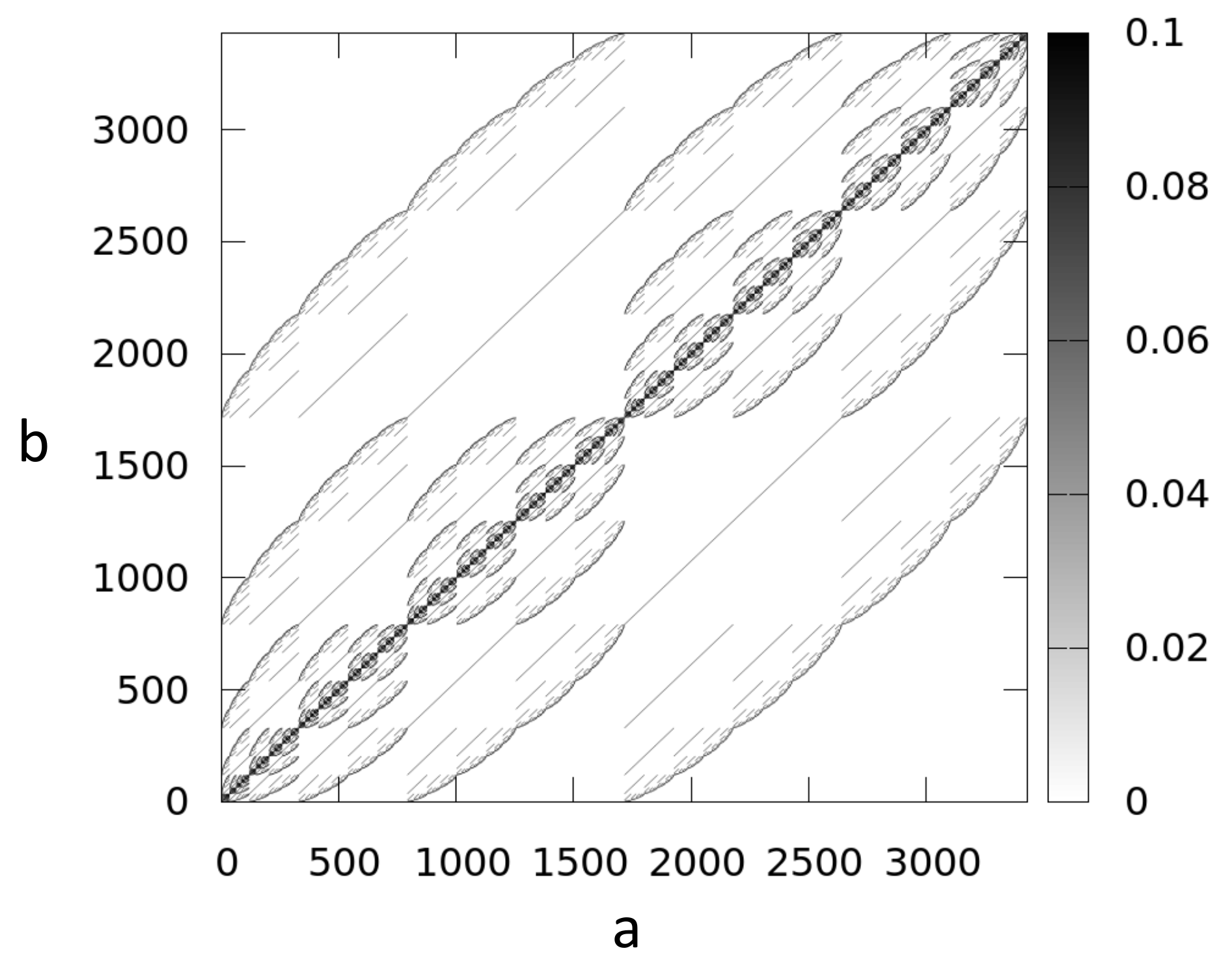}
\caption{Rendering of $|h_{ab}|$ as function of the compressed indices $(a,b)$ for the Hubbard-type model \eqref{Eq:HHHN} with $t_{ij} =e^{- 0.2|i-j|}$, $u_{ij}=0.3 e^{-0.5|i-j|}$ and $\epsilon_i=0$. The plot was generated with \eqref{Eq:CC100}, where the parameters were fixed at $q=14$ and $N=7$, in which case the dimension of the $N$-particle sector was $D_N=3432$.}
\label{Fig:Sparce}
\end{figure}

\begin{code}{\rm We provide here code lines which compute and store the matrix of the Hamiltonian defined in Example~\ref{Hubbard}, this time in the $N$-th particle sector of $\mathcal M_{2^q}$.
\begin{equation}\label{Eq:CC100}
\boxed{
\begin{array}{l}
\mathrm{input} \ N \\
\mbox{Call Eq.~\eqref{Eq:IndN}} \\
h_{a,b}=0, \ a,b = 1,\ldots,D_N \\
\mathrm{do} \ n=0,2^q-1 \\
\ \ \mbox{Call Eq.~\ref{Binary}} \\
\ \  \mathrm{if}(\mathrm{sum}(\mathfrak b_n) =N) \ \mathrm{then}\\ 
\ \ \ \ \mathrm{do} \ i=0,q-1 \\
\ \ \ \ \ \ \mathrm{do} \ j=0,q-1 \\
\ \ \ \ \ \ \ \ \mathrm{if}(\mathfrak b_n(i) =1 \ \& \ \mathfrak b_n(j)= 0)  \ \mathrm{then} \\
\ \ \ \ \ \ \ \ \ \ \mathcal N_{ij}=\mathrm{sum} \big ( \mathfrak b_n(\min(i,j):\max(i,j) ) \big ) \\
\ \ \ \ \ \ \ \ \ \ a=\mathrm{ind}(n); b=\mathrm{ind}(n-2^i+2^j) \\
\ \ \ \ \ \ \ \ \ \ h_{a,b}=h_{a,b}- (-1)^{\mathcal N_{ij}} (1-\delta_{ij})t_{ij} \\
\ \ \ \ \ \ \ \ \ \ h_{b,a}=h_{b,a} - (-1)^{\mathcal N_{ij}} (1-\delta_{ij}) \bar t_{ij} \\
\ \ \ \ \ \ \ \ \mbox{end if}\\
\ \ \ \ \ \ \ \ \mathrm{if}(\mathfrak b_n(i) = \mathfrak b_n(j)= 1)   \ \mathrm{then} \\ 
\ \ \ \ \ \ \ \ \ \ a = \mathrm{ind}(n) \\
\ \ \ \ \ \ \ \ \ \ h_{a,a}=h_{a,a}+\delta_{ij}\epsilon_i + u_{ij} \\
\ \ \ \ \ \ \ \  \mathrm{end \ if}\\
\ \ \ \ \ \ \mathrm{end \ do} \\
\ \ \ \ \mathrm{end \ do} \\
\ \ \mathrm{end \ if} \\
\mathrm{end \ do} \\
{\rm return} \ h_{a,b}, \ a,b = 1,\ldots D_N.
\end{array}
}
\end{equation}
An output of this algorithm is illustrated in Fig.~\ref{Fig:Sparce}.} $\Diamond$
\end{code}

\section{Conclusions}

Although the examples we provided were all 1-dimensional, our analysis covers quite generic settings because, once a basis for the one-particle Hilbert space is chosen, Hamiltonians are all rendered using linear indices. To exemplify this point, let us consider a 2-dimensional $L \times L$ lattice $\mathcal L_L=\mathbb Z_L \times \mathbb Z_L$ ($\mathbb Z_L = \mathbb Z/L\mathbb Z$) with $K$ quantum states per site, as well as a generic Hubbard-type Hamiltonian:
\begin{align}\label{Eq:HLast}
H = \sum_{\bm i,\bm j \in \mathcal L_L} \sum_{\alpha,\beta=0}^{K-1} \Big [ & (1-\delta_{\bm i \bm j} \delta_{\alpha \beta})\big (t_{\bm i \bm j}^{\alpha \beta} \,  a_{\bm i,\alpha}^\ast a_{\bm j,\beta} +\bar t_{\bm i \bm j}^{\alpha \beta} \,  a_{\bm j,\beta}^\ast a_{\bm i,\alpha} \big ) \\ \nonumber 
& \quad  + \big (\delta_{\bm i \bm j} \, \delta_{\alpha \beta} \epsilon_{\bm i}^\alpha+u_{\bm i \bm j}^{\alpha \beta} \big) \, n_{\bm i,\alpha} \, n_{\bm j,\beta} \Big ],
\end{align}
This model can be reduced identically to the Hamiltonian in Eq.~\eqref{Eq:Hub1}, by creating a linear index $i$ for the one-particle Hilbert space $\mathbb C^K \otimes \ell^2(\mathcal L_L)$ of the model. One way to achieve that is by applying the rule:
\begin{equation}
\begin{array}{c}
\mathcal L_L \times \mathbb Z_K \ni (\alpha,\bm i)= (\alpha,i_1, i_2) \\  \Downarrow \\
 i= \alpha + i_1 \, K + i_2 \, L \, K \in \mathbb Z_q,
 \end{array}
\end{equation}
with $q=K\, L^2$. Once we encode the information and re-write the Hamiltonian~\eqref{Eq:HLast} using this linear index, which amounts to re-encoding the coefficients $t_{\bm i \bm j}^{\alpha \beta} \rightarrow t_{ij}$ and $u_{\bm i \bm j}^{\alpha \beta} \rightarrow u_{ij}$, there is nothing to be added to the previous analysis. Of course, not all basis set choices are the same and some can prove to be more optimal, in the sense that the coefficients $t_{ij}$ are of shorter-range. This is an important issue which needs to be solved before the matrix-representation is attempted.

We also want to stress that the calculations can be straightforwardly expanded to cover higher order products of generators. This becomes quite apparent if the reader examines Eq.~\ref{Eq:Z1} and the manipulations after it.  Specific applications taking advantage of these matrix representations will be reported in a future work.

\acknowledgements{This work is supported by National Science Foundation through grant DMR-1823800.}

\bibliographystyle{plain}

\end{document}